# When lost in a multiverse again

Ilya Weinstein

NANOTECH Center, Ural Federal University,

Mira Street, 19, Yekaterinburg, Russia, 620002

i.a.weinstein@urfu.ru

Almost two years ago [1], Andre Geim offered to all travelers found themselves lost in alternate realities to use graphene as a universal metrological tool to quickly determine whether they had come out beyond their Universe. This can be done in two simple ways either by carrying out an optical absorption experiment or estimating the von Klitzing constant in the quantum Hall effect. These measurements using graphene give information about the value of the fine-structure constant $\alpha = e^2/2\varepsilon_0 hc = c\mu_0/2R_K$. The latter reflects the relationship between the fundamental constants: $e$ is the elementary charge, $h$ is the Planck constant, $c$ is the speed of light in vacuum, $R_K$ is the Klitzing constant, $\varepsilon_0$ and $\mu_0$ are the electric and the magnetic constants, respectively. More recently, the metrological role of such experiments has officially increased. The fact is that on May 20, 2019, the International System of Units (SI) accepted some changes and switched to a new global defining set based on exact numerical values of seven constants of nature, including $c$, $e$, and $h$ [2]. In turn, the $\varepsilon_0$ and $\mu_0$ must now be determined experimentally and its standard uncertainty equal to that of the recommended value of the fine-structure constant [2, 3]. It is worth noting that in our Universe $\alpha \approx 1/137 \ll 1$ and such a sighting will be further very useful.

In the Letter, I'd like to dwell on optical measurements in graphene. The works [4, 5] report that the optical transmission $T = 1 - \pi\alpha \approx 97.7\%$ for a graphene monolayer depends only on the fine-structure constant (see the dotted black curve in the Figure). In this case, the absorption increases in proportion to the number of layers. It should be emphasized that the formula is derived accounting for the features of our Universe, namely for the condition $\pi\alpha \ll 1$. A more rigorous expression for the transmission coefficient of graphene layer appears as follows [4, 5]: $T = (1 + \pi\alpha/2)^{-2}$ (see the solid blue line in the Figure). To calibrate an unknown universe, a traveler needs to use this particular curve. The Figure shows that the corresponding dependencies (the dotted black and solid blue lines) begin to noticeably differ when the α increases. For example, for $\alpha = 4/137$, the discrepancy in values reaches $T$ = 90.8 and 91.4%, respectively.

However, there is another essential point to which we are accustomed in our Universe. Nevertheless, it will not necessarily retain its pertinence during voyages over parallel worlds. Here, the matter at hand is the value of the number $\pi$ = 3.14. The latter is determined by the geometry of space and affects the quantitative estimates within many physical laws. The known works [6, 7] on the Minkovski - Banach geometry yield $3 \leq \pi \leq 4$. In this regard, an unexpected metrological benefit of the generalized trigonometry of David Shelupsky [8] cannot fail to be seen. An approach proposed by him implements the relation $\sin^s x + \cos^s x = 1$, where the order $s \geq 1$ is a measure of the curvature of space. Then [8], the generalized number $\pi_s = 2\int_0^1 (1-t^s)^{\frac{1-s}{s}} dt = \frac{2\Gamma\left(\frac{1}{s}\right)^2}{s\Gamma\left(\frac{2}{s}\right)}$ can be expressed in terms of the Gamma functions [9] and takes values in the interval from $\pi_1 = 2$ to $\pi_\infty = 4$. It is easy to verify that for plane Euclidean space $\pi_2 = \pi = 3.14$.

The Figure illustrates the calibration curves for the extreme cases $\pi_1$ and $\pi_\infty$, see the dashed blue lines. Measuring the graphene's optical transparency, a traveler determines the range of values of $\alpha$, and hence a set of universes in which the traveler can currently be. An example of a possible interval for $\alpha$ from $3.7/137$ to $7.5/137$ at $T = 90\%$ is shown with the bidirectional horizontal arrow. On the other hand, if $\alpha = 3/137$, the transmission of the graphene layer in such universes may amount to from 95.6 to 91.2% depending on the generalized number $\pi_s$ (see the bidirectional vertical arrow in the Figure). Consequently, independent information about one of the constants is required for unambiguous identification. The number $\pi_s$ may similarly affect the estimate of $\alpha$ with using measurements of the fundamental resistance $R_K$. This is because in this case we need the value of $\mu_0 = 4\pi \cdot 10^{-7}$ H/m [3, 10]. In conclusion, I'd like to draw attention to the inset in the Figure. It shows the dependence for the transmission of graphene in a wide range of variation of the fine-structure constant with dashed curves for the extremely possible values of the generalized number $\pi_s$. It can be seen that, the sharpest decrease in the $T(\alpha)$ dependence below 90 % is observed when $\alpha > 4.7/137$. For example, 50% of the optical radiation will pass through a graphene layer in a universe where $\alpha = 36.1/137$.

Thus, going for walks through a multiverse and armed with the aforementioned calibration curves, travelers will be able not only find out in which of an infinite number of universes they happened to be, but also get the opportunity to track the relevant curvature of the visited space. To avoid getting lost, they just need to bring along a transparent metrological substrate with graphene. Take the road again!

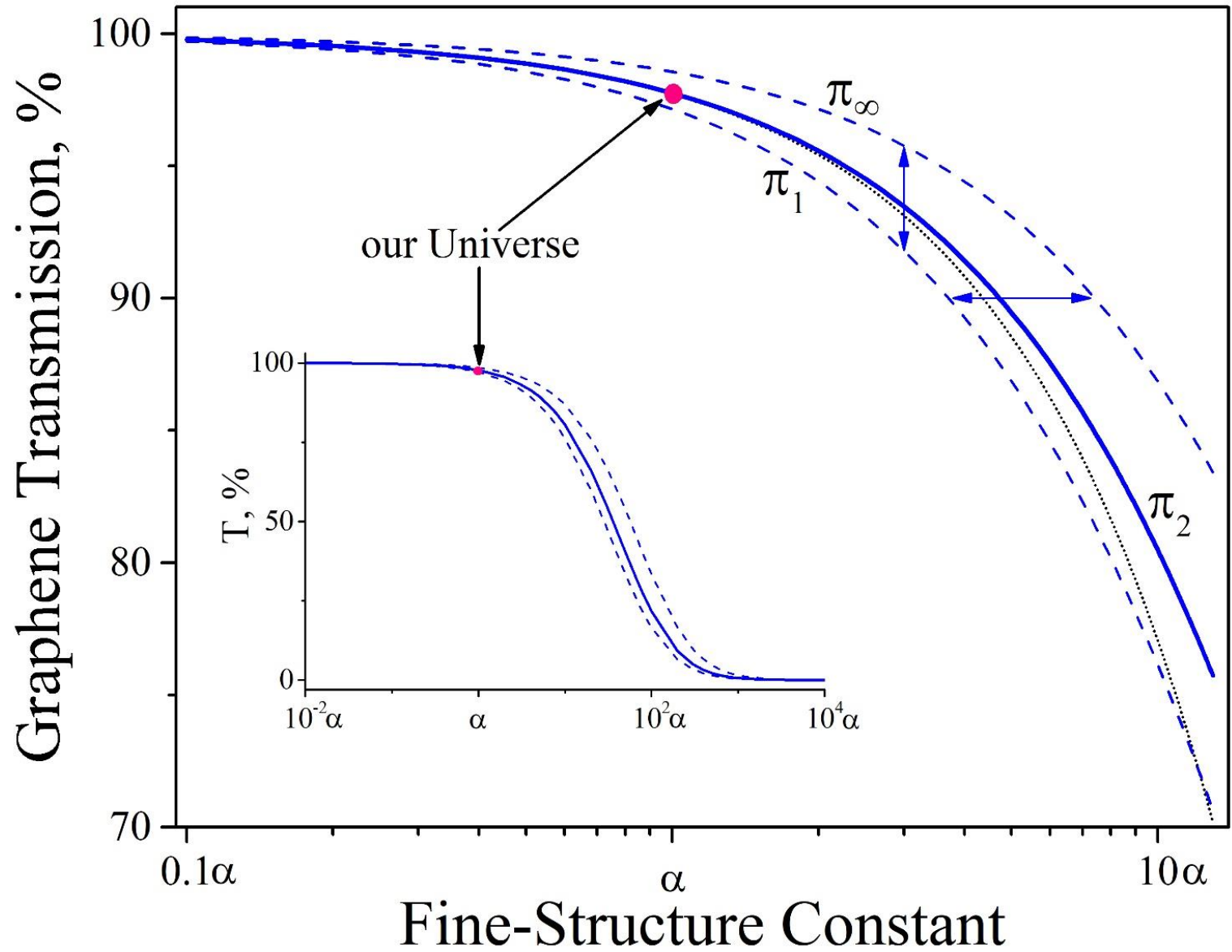

**Figure.** Transmission coefficient of graphene layer for different values of the fine-structure constant and the generalized Pi number.